\documentclass[referee]{raa}
\usepackage{graphicx,times}
\usepackage{natbib}
\usepackage{amssymb,amsmath}
\usepackage{epstopdf}
\usepackage{color}
\usepackage{subfigure}
\usepackage{ulem}
\bibpunct{(}{)}{;}{a}{}{,}

\usepackage{pdflscape}

\usepackage[pagebackref=true]{hyperref}
\hypersetup{pdftitle = The title of my PDF, pdfauthor = My name, pdfsubject= The subject, pdfkeywords = keyword1 keyword2 keyword3}
\hypersetup{colorlinks = true, linkcolor = green, anchorcolor = red, citecolor = blue, filecolor = red, pagecolor = red, urlcolor = red}

\def\cmsq{\hbox{cm$^{-2}$}}
\def\kmps{\hbox{km $\rm{s^{-1}}$}}
\def\nv{N~{\sc v}}
\def\civ{C~{\sc iv}}
\def\siiv{Si~{\sc iv}}
\def\sixiii{Si~{\sc xiii}}
\def\sixiv{Si~{\sc xiv}}

\newcommand\rotate{\@pt@rottrue}%

\def\cmsq{\hbox{cm$^{-2}$}}
\def\kmps{\hbox{km $\rm{s^{-1}}$}}
\def\nv{N~{\sc v}}
\def\civ{C~{\sc iv}}

\begin{document}

   \title{Joint Fit of Warm Absorbers in COS \& HETG Spectra of NGC 3783
}

 \volnopage{ {\bf 2017} Vol.\ {\bf X} No. {\bf XX}, 000--000}
   \setcounter{page}{1}

   \author{Xiao-Dan Fu\inst{1}, Shui-Nai Zhang\inst{1}, Wei Sun\inst{1}, Shu Niu
      \inst{2}, and Li Ji\inst{1}}

   \institute{ Purple Mountain Observatory, Chinese Academy of Sciences, Nanjing 210008,
China; {\it snzhang@pmo.ac.cn}\\
        \and
             Shanghai Astronomical Observatory, Chinese Academy of Sciences,
             Shanghai 200030, China\\
\vs \no
   {\small Received []; accepted []}
}

\abstract{Warm Absorbers (WAs), as an important form of AGN outflows, show absorption in both UV and X-ray band.
Using XSTAR generated photoionization models, for the first time we present a joint fit to the simultaneous observations of {\sl HST}/COS and {\sl Chandra}/HETG on NGC 3783.
Totally five WAs explain well all absorption features from the AGN outflows, which spread a wide range of ionization parameter log$\xi$ from 0.6 to 3.8, column density log$N_H$ from 19.5 to 22.3 \cmsq, velocity $v$ from 380 to 1060 \kmps, and covering factors from 0.33 to 0.75, respectively.
Not all the five WAs are consistent in pressure.
Two of them are likely different parts of the same absorbing gas, and two of the other WAs may be smaller discrete clouds that are blown out from the inner region of the torus at different periods.
The five WAs suggest a total mass outflowing rate within the range of 0.22-4.1 solar mass per year.
}

   \authorrunning{Fu et al.}            %author_head in even pages
   \titlerunning{Warm Absorbers in NGC 3783}  % title_head in odd pages
   \maketitle

%________________________________________________ sections below
%
\section{Introduction}           %% first-level sections will be auto-capitalized
\label{sect:intro}

Warm absorbers (WAs), known as an important form of Active Galactic Nuclei (AGN) feedback, are likely having an effective impact on the host galaxy, and even the intergalactic environment.
WAs, the outflowing ionized gas that generates absorption in the X-ray and UV bands,
are detected in about 50$\%$ Type I Seyfert galaxies either in UV (\citealt{Crenshaw+etal+1999}) or in X-ray band (\citealt{Reynolds+etal+1997}).
It offers great opportunities to study the specific processes of gas around black holes such as the angular momentum transfer during accretion, and also how the AGN feedback affects host galaxies.

A lot of efforts have been spent to study the WAs in the past three decades (since identified by \citealt{Halpern84}), but nevertheless the WA properties and origins are not fully understood.
Nowadays the strategy for WA studies is to focus on a few exceptional Seyfert galaxies, and cumulate a large amount of data, many of which are simultaneous multi-wavelength spectral observations.
UV spectrum has higher spectral resolution and accurate dynamic measurements, corresponds to a small number of atomic transitions.
In contrast, the X-ray spectrum covers more transitions from ions at various ionization states, but is subject to lower resolution and observation precision (e.g., \citealt{Kaspi+etal+2002}).
Combination of High-resolution UV or X-ray spectroscopy brings more comprehensive WA studies (e.g., \citealt{Costantini+2010}).
Tens of joint UV-X-ray campaigns have provided much insight into the physical conditions of these absorbing outflows.

However, a broad understanding of these WAs in the two bands has yet to emerge.
One important reason is that the ionization levels and kinematics of both the UV absorbers and X-ray WAs are measured through different methods, and frequently there is only partial overlap in the conditions of the two sets of absorbers (\citealt{Crenshaw+Kraemer+George+2003}).
\citeauthor{Zhang+etal+2015} (2015, hereafter as Z15) upgraded XSTAR\footnote{https://heasarc.gsfc.nasa.gov/lheasoft/xstar/xstar.html} and provided a method to fit every physical components in UV spectra the same way used in fitting X-ray spectra.
As a consequence, the observed spectra can be fitted simultaneously, providing better constrains on the properties of WAs based on the combined information from the UV and X-ray regimes.

Seyfert I galaxy NGC 3783 ($\sim$41.6 Mpc, from NED\footnote{http://ned.ipac.caltech.edu}) is a perfect target as it is very bright in the X-ray and UV bands, and exhibits strong absorption features detected with many instruments such as {\sl ROSAT, ASCA, Chandra, XMM-Newton, Suzaku, FUSE, HST}/STIS, and {\sl HST}/COS (e.g., \citealt{Turner93, Shields97, Kaspi+etal+2002, Gabel+etal+2003b, Bluestin+etal+2002, Netzer+etal+2003, Krongold+etal+2003, Brenneman11}).
The accumulative exposure time on its nuclear region taken by {\sl Chandra}/HETG is more than 1 Ms.
Simultaneous UV-X-ray observations of NGC 3783 was taken in both 2001 and 2013, which show clear variations in velocity of WAs (\citealt{Scott+etal+2014}).
However, the derived physical parameters such as column density $N_H$, ionization parameter $\xi$, and velocity $v$, vary from authors to authors among the X-ray studies,
while the UV regime is more promising to give precise $v$  but less constrained $N_{\rm{H}}$ and $\xi$ from individual lines (discussed in Section 5.1).

A major obstacle in the WA studies is that the contribution of WAs identified in the UV band is not well accounted in the X-ray spectra, vice versa.
In this paper, we aim to identify WAs from the simultaneous HETG and COS observations of NGC 3783 in 2013, using the joint fitting method presented in Z15.
Section 2 describes the observations and data reduction.
In Section 3 we analyze the radiation and absorption components in the spectra and prepare the suitable models for them.
We give the fitting results in Section 4. The discussion and exploration are presented in Section 5.
We end with a summary of our conclusions.

% Authors can give a citation as `\citealt{Michel+etal+1992}'.
% You may also use \cite, \citep and \citet for citation, and use Table~1
% or Figure~1 and so forth. Using \ref and \label for cross-references of
% Tables/Figures is a good way in adjusting/adding/removing text, tables or
% figures.

\section{Observation and data reduction}
\label{sect:Obs}

NGC 3783 was observed by {\sl Chandra}/HETG and {\sl HST}/COS simultaneously in March 2013 (PI: William Brandt) with exposure times of 160 ks and 4 ks, respectively.
The observations were closely conducted within five days.

\subsection{COS Spectrum}
The COS spectrum covers the wavelength band of 1135-1795 \AA~(\citealt{Osterman+etal+2011}), consisting of G130M and G160M data (Grating centered at 130 and 160 nm) where M means the medium spectral resolution $R\equiv\lambda/\Delta\lambda$ from 16000 to 21000.
In fact, this NGC 3783 observation in 2013 consists of eight discrete shorter observations, four from G130M and four from G160M respectively, which are all primarily calibrated and processed through the pipeline CAL$\mathit{COS}$ before being downloaded from Multi-mission Archive for Space Telescopes (MAST\footnote{http://archive.stsci.edu}).
Flat-fielding, alignment, and co-addition of the processed exposures are carried out using IDL routines described in \cite{Danforth+etal+2010}.
They are merged with exposure weighting, and the final spectrum has signal-to-noise (S/N) ratios per resolution element (0.07 \AA, or 17 \kmps) ranging from 15-25.
Finally the COS flux spectrum is converted into the format commonly used in the X-ray studies through the IDL tool PINTOFALE\footnote{ http://hea-www.harvard.edu/PINTofALE/}.
A Pulse Height Amplitude (PHA) file is obtained, with the corresponding response (RSP) file that convolves G130M and G160M line spread functions (LSF).
Figure 1 presents the COS spectrum of NGC 3783.

\begin{figure}
\centering
\includegraphics[angle=0,width=6in]{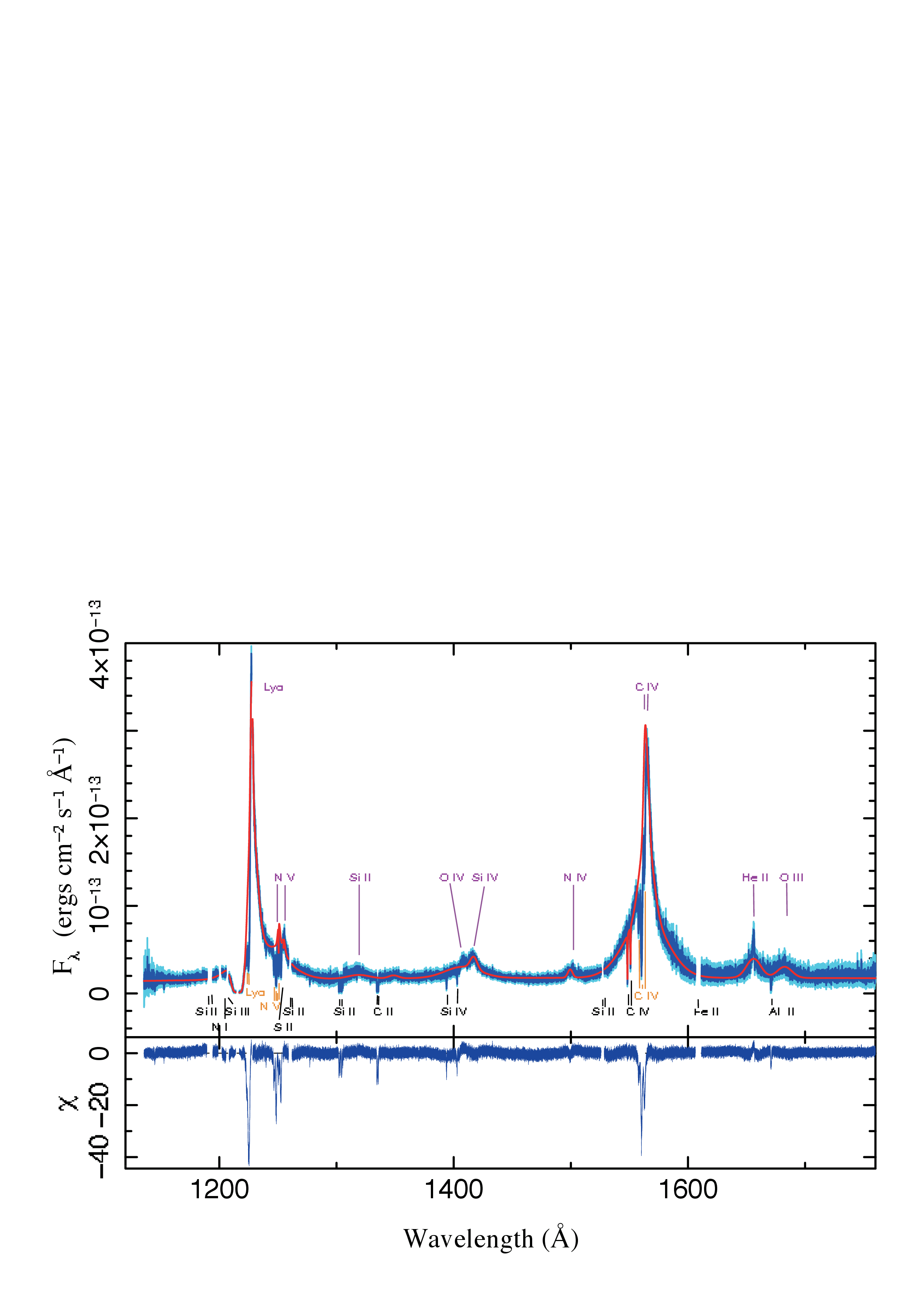}
\caption{The COS spectrum of NGC 3783, where the data is in blue and the error bar is in cyan. Emission lines are labeled with purple notes, and absorption lines from NGC 3783 and local ISM are with orange and black notes, respectively.  The red curve shows the best-fit model of continuum and line emission.}
\end{figure}

\subsection{HETG Spectrum}
As joint observations, {\sl Chandra} observed NGC 3783 with HETG instrument twice within two days in March 2013, which have exposure time of about 60 ks (ID: 14991) and 100 ks (ID: 15626), respectively.
The HETG consists of two sets of gratings: the Medium Energy Grating (MEG) which covering the wavelength range of 2.5-31 \AA~and the High Energy Gratings (HEG) with the range of 1.2-15 \AA.
Since NGC 3783 is heavily obscured by WAs especially in the longer wavelength band, only 2-11 \AA~band is taken into account in this work.
The two observations are not apparently contaminated by other sources in the field of view, but the zeroth order images are confronted with the pile-up issue.
Therefore we run the ISIS script $\rm TG\_FINDZO$\footnote{http://space.mit.edu/cxc/analysis/findzo/index.html} to calculate a more accurate position of this source and use the standard data reduction steps of {\sl Chandra} Interactive Analysis of Observations\footnote{http://cxc.cfa.harvard.edu/ciao/threads/diffuse emission} (version 4.9) to get the spectra and corresponding response (RMF and ARF) files.
For the spectral analysis, we only use the first order spectra for the better S/N ratio.
Figure 2 is the combined MEG and HEG spectrum of NGC 3783.

\begin{figure}
\centering
\includegraphics[angle=-90,width=6in]{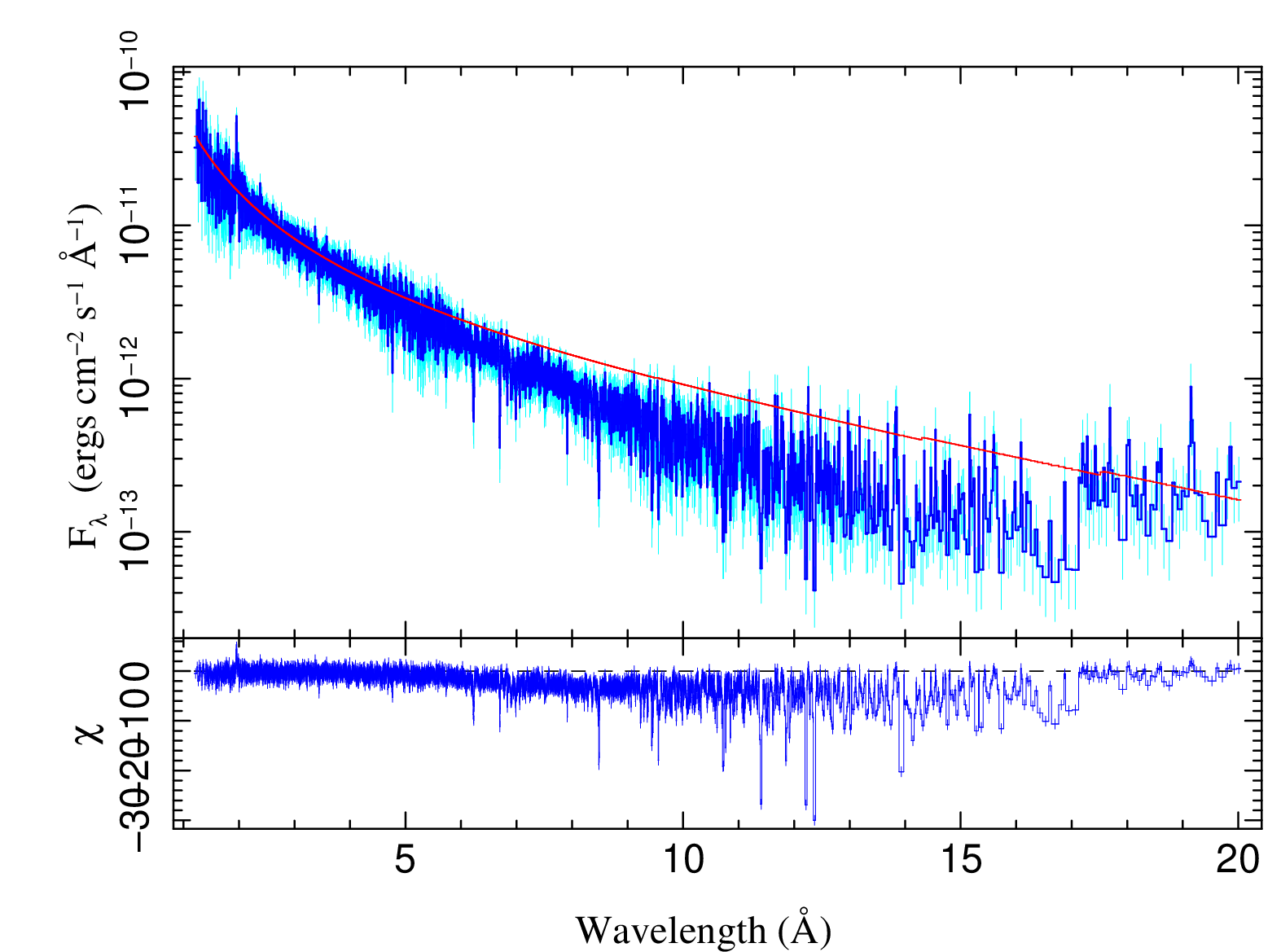}
\caption{The Combined HETG spectrum. The residual panel shows the absorption from WAs.}
\end{figure}

\section{Analysis and Model preparation}
We use the Interactive Spectral Interpretation System (ISIS\footnote{http://space.mit.edu/cxc/isis/}; version 1.6.2, \citealt{Houck02}) to analyze and fit the spectra.
ISIS is a programmable, interactive tool for studying the physics of spectra.
The emission and absorption lines are identified based on AtomDB\footnote{http://www.atomdb.org} (version 3.0.3) (\citealt{Foster+etal+2012}).

The photoionization code XSTAR (Version 2.2, \citealt{Kallman01}) is employed to model the emitting and absorbing plasma that are photoionized by the central AGN.
In XSTAR models, the intrinsic free parameters are column density $N_H$ and the ionization parameter $\xi =L_{ion}/(nr^{2})$, where $L_{ion}$ is the luminosity in the 1-1000 Ryd energy range, $n$ is the hydrogen density, and $r$ is the distance to the ionizing source.
The metal abundances are generally set to solar values.

\subsection{Local Absorption in Milky Way}
The column density of Galactic gas absorption is about $10^{21}$ \cmsq~along the line of sight to NGC 3783 (\citealt{Kalberla+etal+2005}), which is included in all the later analysis by employing the ISM absorption model TBnew\footnote{http://pulsar.sternwarte.uni-erlangen.de/wilms/research/tbabs/}.
Dust extinction can effectively reduce the UV radiation.
We take this effect into consideration by using the extinction curve formula proposed by \cite{Gordon+2009},
%, which covers the wavelength range from 910 $\rm{\AA}$ to 3.3 $\mu$m.
with parameter ${R_{\mathrm{v}} =A(V)/E(B-V)=3.1}$ (\citealt{Cardelli89}), and ${A(V)=0.332}$ in the case of NGC 3783.
%No free parameter is used in the dust extinction component.

\subsection{Spectral Energy Distribution of NGC 3783}
It is necessary to construct the spectral energy distribution (SED) of NGC 3783 for generating the photoionization models.
The HETG and COS spectra are put into one plot (Figure 3).
The intrinsic UV continuum from the accretion disk can be corrected from the local absorption.
The ionizing spectrum in X-ray band is obtained by fitting the continuum in the 2-6 keV band which is not severely affected by WAs.
The local absorption is strong between the HETG and COS bands, as shown by the red line, thus no more data points are likely to be found in between.
We use the typical AGN SED (\citealt{Elvis94}) but scaled with wavelength ($\propto \lambda^{-0.27}$) to match the intrinsic UV and X-ray continuum, in order to mimic the intrinsic SED of NGC 3783.
The ionizing luminosity is then obtained as $L_{ion}=7.6\times10^{43}\,{\rm erg\,s^{-1}}$.

\begin{figure}
\centering
\includegraphics[angle=-90,width=6in]{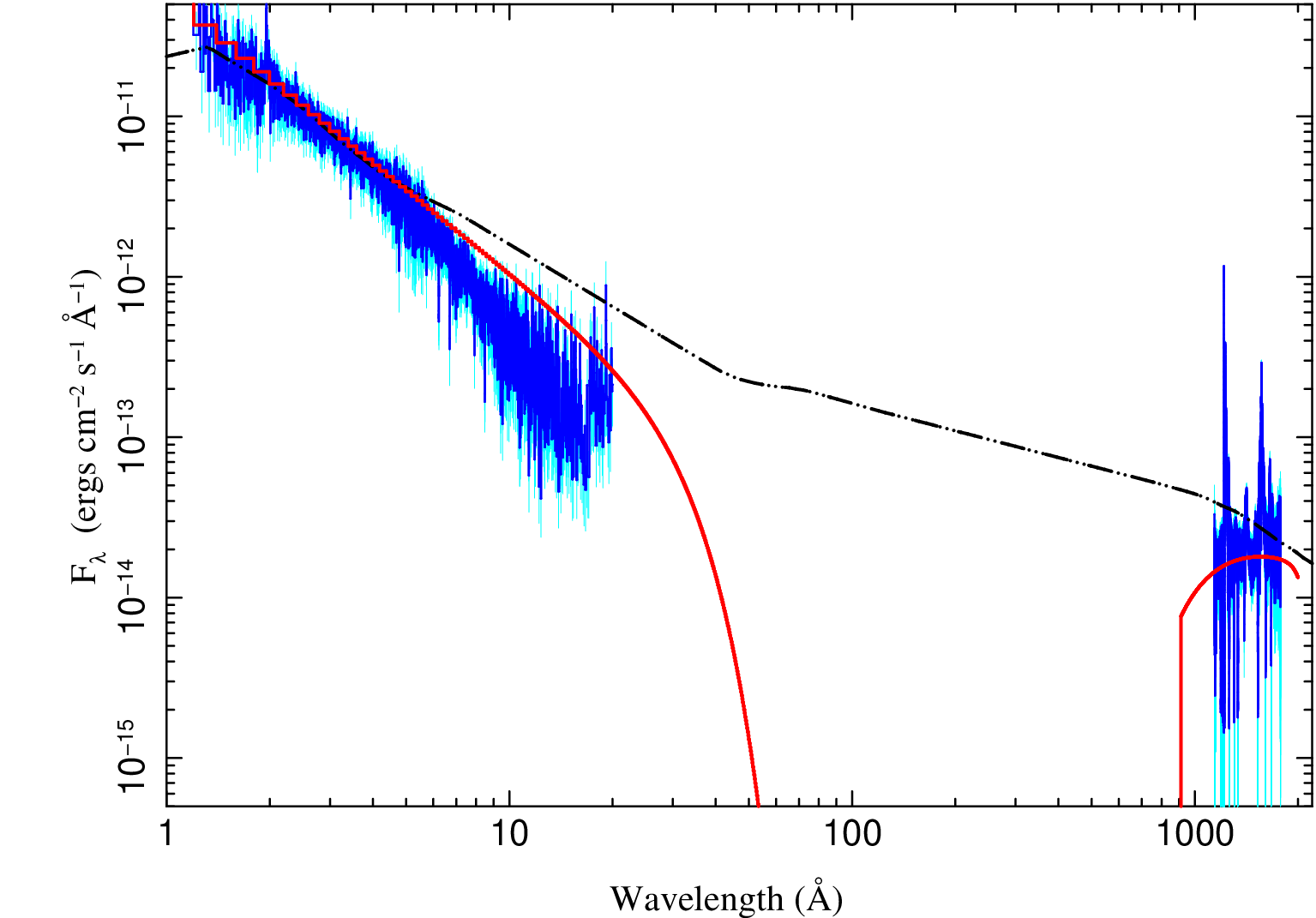}
\caption{The HETGS and COS spectra to construct the SED. The red line shows the hot coronal and accretion disk radiation model after the Milky Way's gas absorption and dust extinction. The black dash-dotted line is the scaled average AGN SED (\citealt{Elvis94}) that approaches the SED of ionizing source in NGC 3783.}
\end{figure}

\subsection{Emission in NGC3783}
The intrinsic radiation in NGC 3783 has different origins in the UV and X-ray bands.
In the X-ray band, the continuum radiation mostly comes from the Comptonized corona at the inner region of the accretion disk.
We use a single power-law model to account for it.

The intrinsic UV radiation on the other hand generally comes from the accretion disk, board line region (BLR), and narrow line region (NLR).
The multiple blackbody emission in the accretion disk can also be fitted empirically by a power-law model.
The radiation from photoionized clouds in BLR and NLR however needs to be generated by XSTAR.

As in Z15, we take the plausible assumption that the clouds in BLR have approximately virial speed, i.e. $v\propto \sqrt{{\rm G}M/R}$, where
$M$ represents the black hole mass and $R$ represents the distance to the center.
In this case, in the BLR are several groups of clouds with different velocities.
Each group is assumed having the same ionization condition and the same full width at half-maximum (FWHM) value,
thus taken as one photoionization component.
The clouds in the NLR have much smaller rotational velocity ($<$2000 km s$^{-1}$), which are also described by one photoionization component.

To obtain the line-broading paramter $b$ ($\rm{FWHM}=2\sqrt{\rm{ln}2} b$) for generating those photoionization components in XSTAR, we fit strong emission lines with Gaussians.
The best target for profile decomposition is the Ly$\alpha$, \nv~doublet, and \civ~doublet, which have high S/N ratio in the COS spectra.
We fit the five lines jointly according to their rest wavelength relations, taking the strong and weak line flux ratio of the doublets as in an optically thin case 2:1.
Three groups of Gaussians with FWHM values of $\sim$8100, 2500, and 600 \kmps~are needed (Table 1), in which the former two indicate the virial velocities in the BLR and the latter is for the NLR.
The redshift of the NLR components $z$=0.009815(19), slightly larger than the $z$=0.009760 of the host galaxy obtained by \cite{de+1991}, will be used as the systemic redshift in this work.

We generate two XSTAR table models for the BLR and one table model for the NLR.
The densities of the BLR and NLR are set to $10^{10}$ cm$^{-3}$ and $10^3$ cm$^{-3}$, respectively (Z15).
For both cases, the column density is set to $10^{23}$ \cmsq, the temperature is set to 15 000 K, the metallicities are set to the solar values, while the ionization parameter and the redshift are left as free parameters.

\subsection{Absorption in NGC3783}

The HETG spectrum, which is grouped to 0.01 \AA~per bin, reveals tens of absorption lines, mostly from k-shell transitions of H-like and He-like ions of O, Ne, Mg, Si, and S and L-shell transitions of Si and Fe (Figure 2).
The L-shell Si lines are from lower ionized absorbing gas, while the two strong lines \sixiii~at 6.7 \AA~and \sixiv~at 6.2 \AA~suggest higher ionization.
For the highly ionized WAs, we take the line-broadening parameter $b=\sqrt{v_{\rm th}^2+2v_{\rm turb}^2}$ of 350 \kmps~estimated from the 900 ks HETG observations in 2001 (\citealt{Kaspi+etal+2002}), which is larger than MEG spectral resolution of $\sim$170 \kmps.

The outflow velocities of WAs can be better estimated from the \nv~and \civ~doublet troughs in the COS spectrum.
By fitting with Gaussians, we find dynamic components with velocities of 600, 880, and 1070 \kmps~(Table 1),  of which the second one is identified by the absorption dip of N V 1239 \AA~line, and the asymmetric shape of the C IV 1548 \AA~absorption line.
To determine the turbulent velocity $v_{\rm turb}$ is s somewhat complex since the thermal broadening can make an important contribution.
According to $\rm{FWHM}=2\sqrt{\rm ln2}\sqrt{v_{\rm th}^2+2v_{\rm turb}^2}$, turbulent velocities are obtained from their fitted FWHM.
$v_{\rm th}$ represents the thermal velocity, which is about $v_{\rm th}=13\sqrt{T_4/A}$ where $T_4$ is the temperature in unit of $10^4$ K and $A$ is the atomic number.
The temperatures of these lower ionized WAs are about $10^{4.4}$ K according to the thermal stability curve of NGC 3783 generated by \cite{Netzer+etal+2003}, and
meanwhile, we take the mean value of carbon and nitrogen atomic numbers as the $A$ value.
All the turbulent velocity $v_{\rm turb}$ values and outflowing velocities are listed in Table 1.

Based upon the lack of response of WAs to the changes of continuum, some lower distance limits are given (\citealt{Netzer+etal+2003, Behar+etal+2003}).
They may be located between the inner region of torus to the NLR ($\sim$1-25 pc), as assumed in our model.
While the WAs can easily cover the region of accretion disk, they usually do not cover the whole BLR that has a transverse size of $1.9\times10^{16}$ cm (\citealt{Onken02}).
We use the XSPEC model `partcov' to mimic this effect when fitting the COS spectrum.

At last, we generate XSTAR models for these WAs.
The column density $N_{\rm H}$, the ionization parameter $\xi$, and the redshift $z$ are the free parameters.
The WA with velocity of 600 \kmps~is saturated in the \nv~and \civ~absorption lines, which has been taken into account by XSTAR.

\begin{table*}
\begin{center}
\caption{Gaussian fitted intrinsic lines in COS spectrum.}
\small
\begin{tabular}{ccc|cc|cc} \hline\hline
$ Ion (\lambda_{\textsl{rest}})$   & $f$ & $flux\,(\times10^{-4})$   &   $\lambda_{\textsl{obs}} $      & $Velocity$   & $\rm{FWHM}$    & $v_{\rm turb}$  \\
$\textrm{\AA}$    &           &$\mathrm{photons^{-1}~cm^{-2}}$  &$\textrm{\AA}$ & $\kmps $ & $\kmps$  &$ \kmps$ \\
\hline\noalign{\smallskip}
\multicolumn{3}{c}{ Emission lines from BLR and NLR} \\
\hline\noalign{\smallskip}
Ly$\alpha$ (1215.67)  &0.41617   &2606.3  &$1227.57 $    &        &$8101$	        & 3440  \\
                      &	         &2882.2 &$1227.73$     &        &$2514$	        & 1068  \\
                      &	         &771.4   &$1227.60$     &        &$596$ 	        &253  \\
\nv~(1238.82)  & 0.15553  &728.8    &1250.95      &        &$8101$	        &3440 \\
                      &	         &176.8    &1251.11	     &        &$2514$	       &1068 \\
                      &	         &121.9    &1250.98      &       &$595$ 	        &253  \\
\nv~(1242.81)  & 0.077805 &364.4     &1254.98     &       &$8101$	       &3440 \\
                      &	          &88.4     &1255.14     &       &$2514$	        & 1068 \\
                      &	         &60.9      &1255.01     &       &$596$ 	        &253 \\
\civ~(1548.19)   &0.19045   &3752.3   &$1563.35\pm0.12$ & $7\pm23$ &$8101\pm77$    & $3440\pm33$ \\
                       &          &1627.5   &$1563.55\pm0.08$ & $-32\pm16$ &$2514\pm36$  &$1068\pm16$ \\
                      &           &682.6    &$1563.39\pm0.03$ &  (set) $0\pm6$ &$596\pm18$  &$253\pm8$ \\
\civ~(1550.78)  & 0.094824  &1876.1   &$1565.97$   &       &$8101$	  & 3440   \\
                      &           &813.8    &$1566.17$   &       &$2514$      & 1068   \\
                      &           &341.3    &$1566.00$   &       &$596$       & 253   \\
\hline\noalign{\smallskip}
\multicolumn{3}{c}{ Absorption lines from Warm Absobers} \\
\hline\noalign{\smallskip}
\nv~(1238.82)   &0.15553  &$116.7\pm3.8$&$1248.50$ &600      &$198$	       &83       \\
                       &	     &$46.8\pm1.3$ &$1246.57\pm0.03$ & $1067\pm5$&$192\pm17$  &$81\pm7$  \\
                       &        &$26.9\pm1.3$ &$1247.34\pm0.03$ & $880\pm8$    &$156\pm15$ &$65\pm7$\\
\nv~(1242.81)   &0.077805 &$98.3\pm3.8$ &1252.52	&600    &198	     &83         \\
                       &         &$37.5\pm1.3$ &1250.58     &1067   &192	    &81           \\
                       &	     &$27.7\pm1.8$&1251.36     &880    &156	    &65          \\
\civ~(1548.19)   & 0.19045  &$374.3\pm2.2$ &1560.28    & 600    &198	      &83        \\
                       &	      &$114.7\pm3.5$&1557.87    &1067    &192	      &81        \\
                       &	      &$56.7\pm3.7$ &1558.84   &880     &156	     &65         \\
\civ~(1550.78)   &0.094824 &$282.3\pm14.9$ &$1562.90\pm0.01$ &$600\pm2$ &$198\pm19$ &$83\pm8$\\
                      &	         &$91.7\pm3.8$ &1560.48    &1067    &192	     &81         \\
                      &	         &$90.5\pm4.6$  &1561.45   &880     &156	     &65         \\
\hline\noalign{\smallskip}
\end{tabular}
\end{center}
\end{table*}

\section{Joint Fitting and Five Warm Absorbers}

The HETG and COS spectra are well fitted by these physical components.
The photon indices of power-laws in the X-ray and UV bands are $\Gamma=1.07\pm0.02$ and 2.43$\pm$0.04 respectively, as shown by the red line in Figure 3.
The XSTAR models for BLR and NLR can nicely reproduce the COS emission lines except for the \siiv~line at 1416 \AA~that needs an additional Gaussian component (Figure 1).
A Gaussian absorption model is used to account for the local broad Ly$\alpha$ absorption.

Totally five WAs are used to fit the absorption features in both the COS and HETG spectra (Figure 5 \& 6).
The parameters of the best-fit WAs are listed in Table 2, and their synthetic spectral models in the two bands are shown in Figure 4.
WA 1-4 have absorptions in both the UV and X-ray bands, while WA 5 only appears in the X-ray band.
WA 1 \& 2 have similar low ionization states, similar column densities, and their velocities are higher than other WAs.
WA 3 \& 4 have similar velocities and column densities, but at slightly different ionization states.
Now that WA 1-4 are nicely constrained in the HETG spectrum, WA 5 seems highly ionized, with the highest column density.

\begin{figure}
\centering
\includegraphics[angle=-90,width=5.4in]{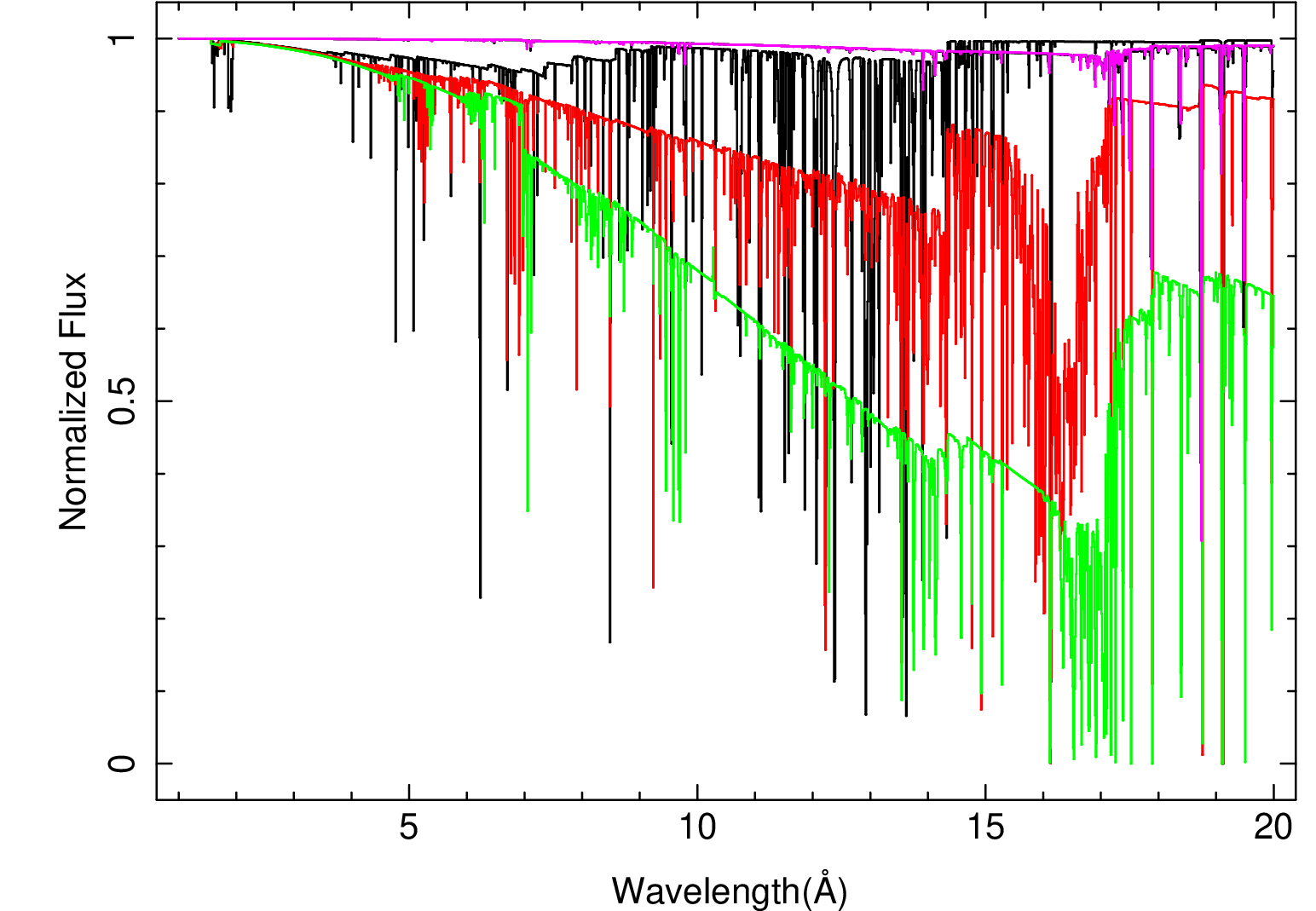}
\\~~\\~~\\
\includegraphics[angle=-90,width=5.4in]{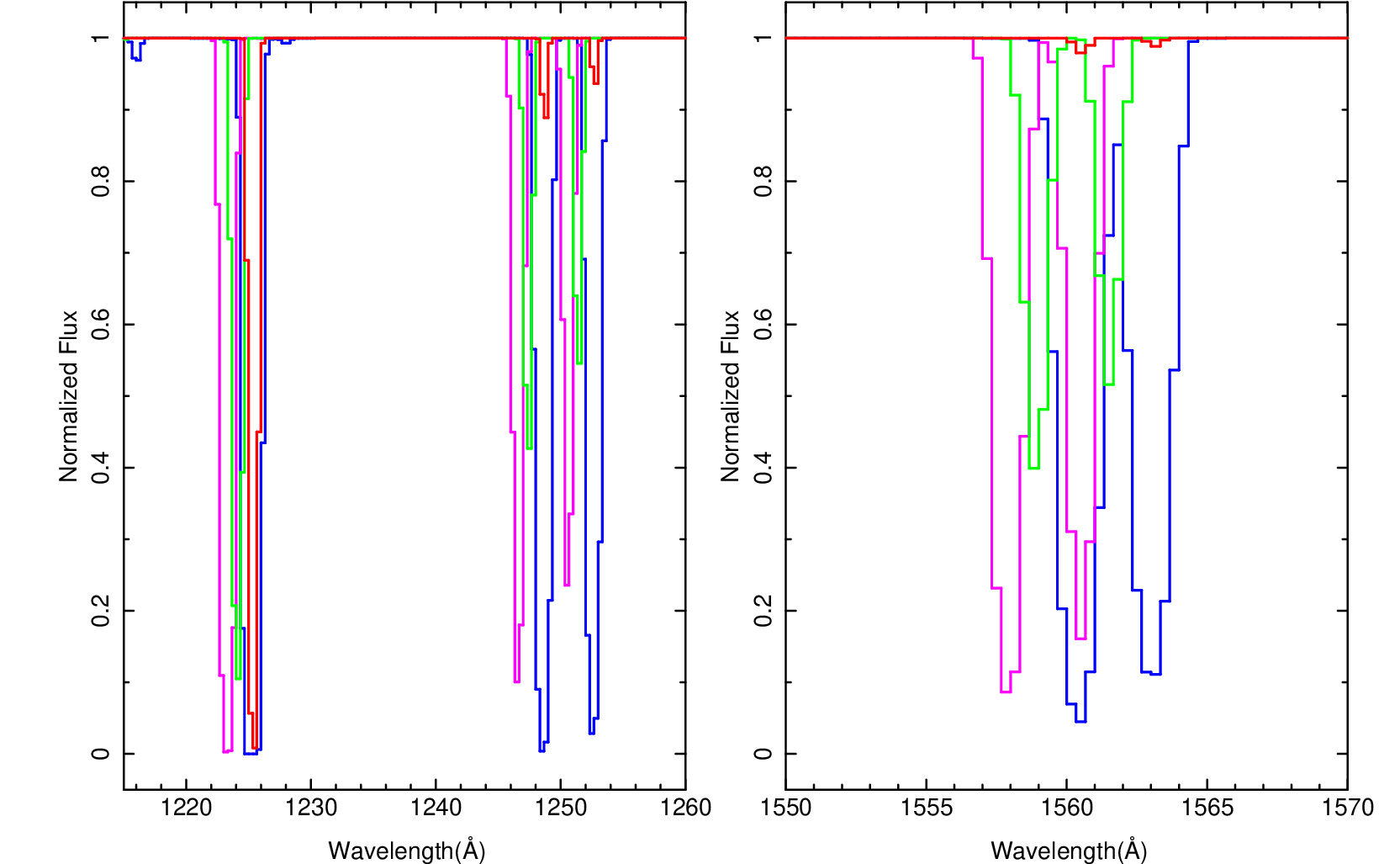}
\caption{The separate models of the WAs: WA1-blue, WA2-magenta, WA3-green,  WA4-red, and WA5-black.}
\end{figure}

\begin{figure}
\centering
\includegraphics[angle=-90,width=2.7in]{ms0095fig5a.eps}
\includegraphics[angle=-90,width=2.7in]{ms0095fig5b.eps}
\caption{The best-fit XSTAR models for Ly$\alpha$, \nv~and \civ~absorptions in NGC 3783. The positions of WA1-4 are labelled with arrows. The absorption lines in 1227.89, 1253.82, 1548.26 and 1550.85 \AA~are from local ISM.}
\end{figure}

\begin{figure}
\centering
\includegraphics[angle=0,width=6in]{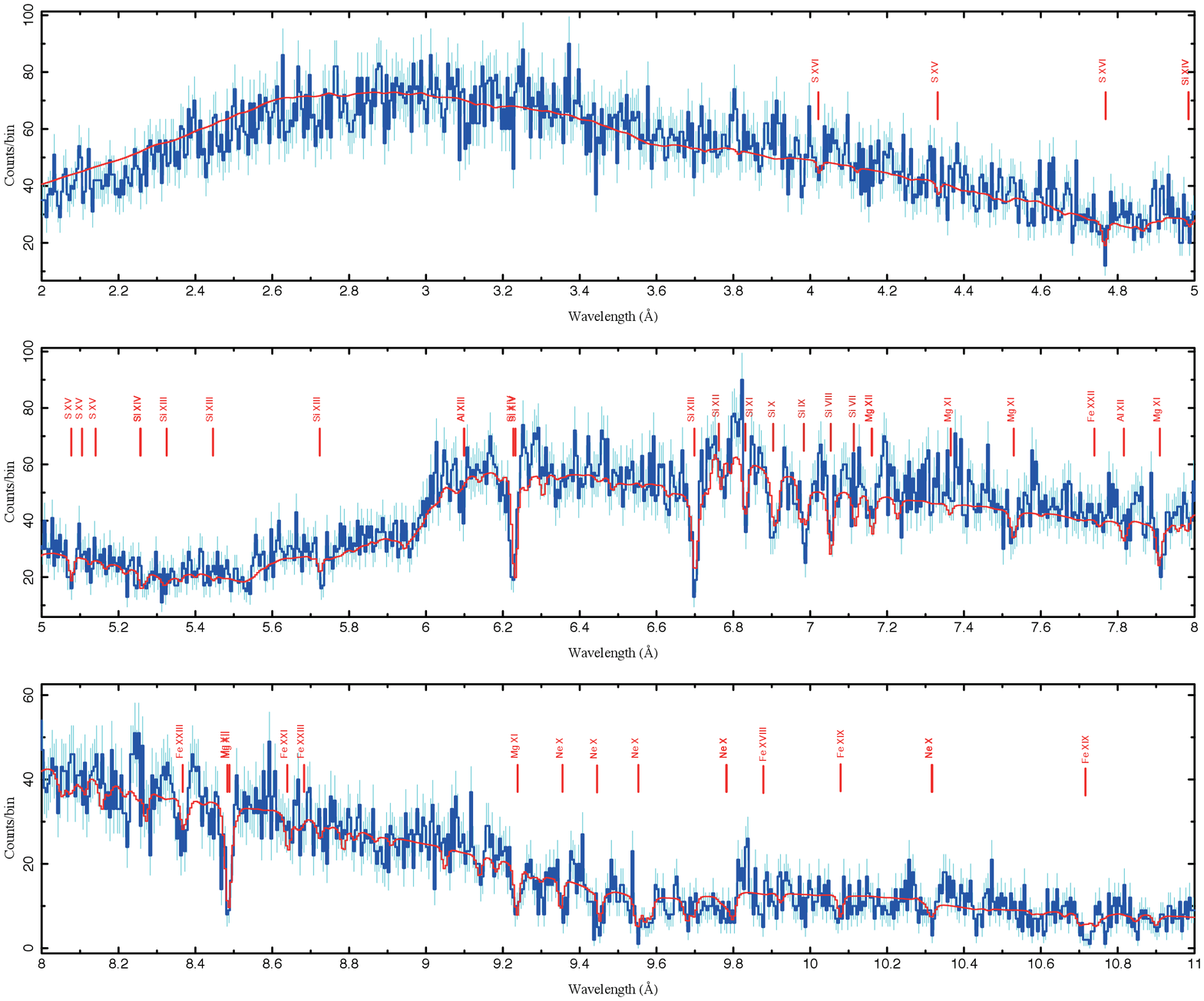}
\caption{The HETG spectrum in blue and the best-fit model in red.}
\end{figure}

\begin{table}
\begin{center}
\caption{Parameters of Warm Absorbers}
\footnotesize
\begin{tabular}{ccccc}
\hline\noalign{\smallskip}
   & log$\xi$       & log$N_{\rm{H}}$               & Outflow velocity     &Covering \\
               &$\rm{erg~s^{-1}~cm}$  &$\rm{cm^{-2}}$       &km s$^{-1}$    &factor \\
\hline\noalign{\smallskip}
\hline\noalign{\smallskip}
WA1       &$0.58\pm 0.03$  &$19.88\pm0.05$    &$1060\pm3$   & $0.33\pm0.01$ \\
WA2       &$0.76\pm 0.07$  &$19.49\pm0.08$      &$870\pm4$   & $0.38\pm0.01$\\
WA3	      &$1.28\pm0.01$   &$21.76\pm0.02$  &$577\pm2$   & $0.75\pm0.01$ \\
WA4       &$2.16\pm0.07$   & $21.76\pm0.04$  &  $550\pm12$   &  $0.40\pm0.03$ \\
WA5       &$3.78\pm0.13$  & $22.27\pm0.03$  &   $380\pm29$  &  - \\
\noalign{\smallskip}\hline
\end{tabular}
\end{center}
\end{table}

\section{Discussion}
\label{sect:discussion}
\subsection{Comparison of WAs with Previous Studies}

\begin{figure}
\centering
\includegraphics[angle=0,width=3.5in]{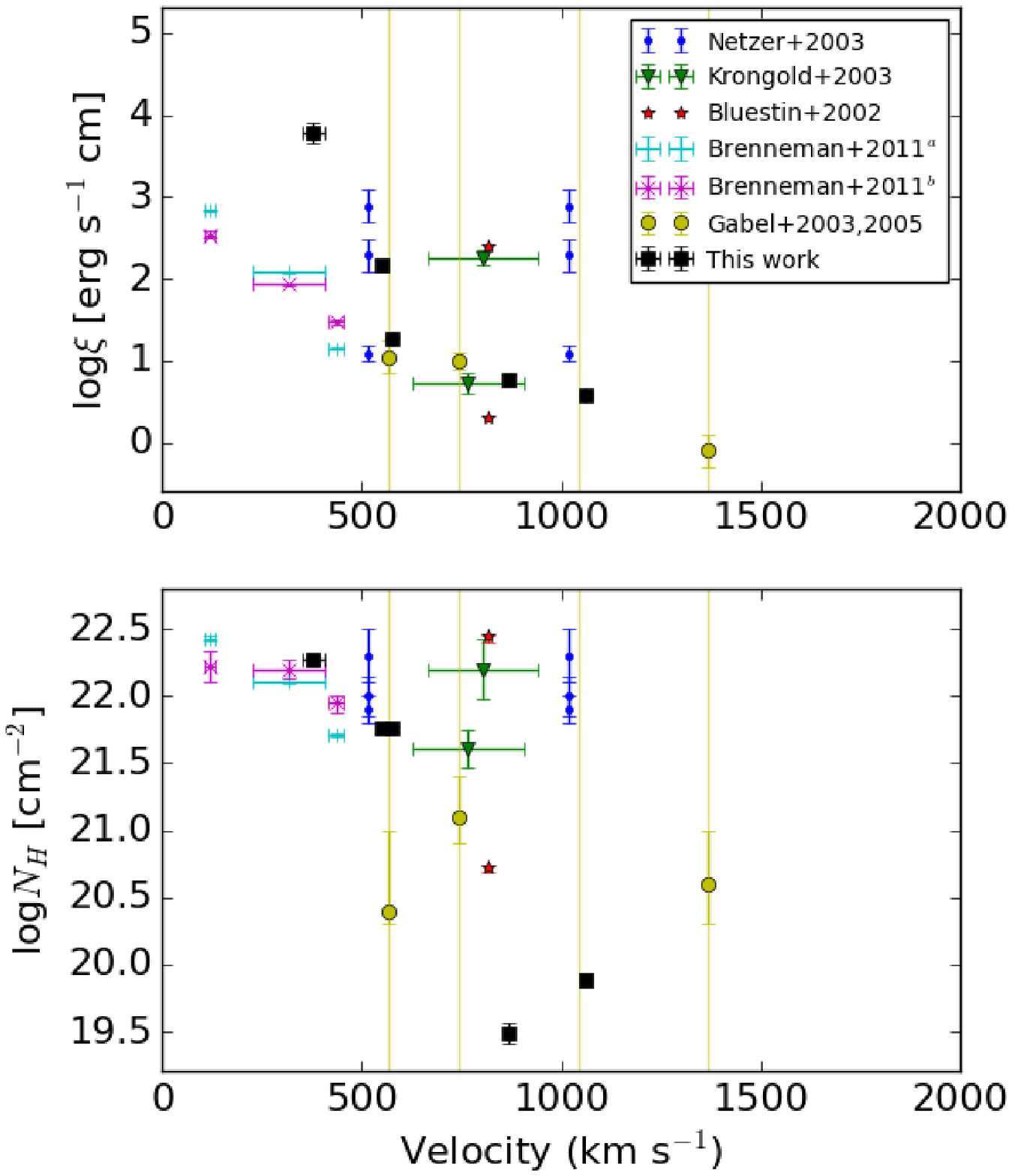}
\caption{Comparison of WA parameters with previous works. The five WAs this work are marked as black squares, where from right to left are WA 1 to WA 5, respectively. \cite{Netzer+etal+2003}, \cite{Krongold+etal+2003}, and \cite{Brenneman11}a use HETG observations in 2001. \cite{Bluestin+etal+2002} use RGS data. \cite{Brenneman11}b use Suzaku data. \cite{Gabel+etal+2003b} use STIS and {\sl FUSE} data to determine the velocities of WAs as marked with the four golden vertical lines. \cite{Gabel+etal+2005} constrain the $N_{\rm H}$ and $\xi$ for three of the four WAs.}
\end{figure}

The WAs in NGC 3783 have been extensively studied in both the UV and X-ray bands even before the high-resolution grating observations (e.g., \citealt{Turner93, Shields97}).
Later on, based on the high resolution spectra, different photoionization models are applied to describe the WAs (as summarized in \citealt{Scott+etal+2014}).
These models usually have the same column density parameter $N_{\rm H}$, but different ionization parameters such as $U$, $U_{OX}$, or $\xi$.
In order to apply the comparison among different studies, we convert them all to $\xi$ as used in XSTAR by: ${\rm log}U={\rm log}\xi-1.5$ (\citealt{Crensha+Kraemer+2012}) and
${\rm log}U={\rm log}U_{\rm OX}+1.99$ (\citealt{Krongold+etal+2003}).
The spectral modeling results are mainly from the X-ray spectra: {\sl Chandra}/HETG observations (\citealt{Netzer+etal+2003, Krongold+etal+2003, Brenneman11}), {\sl XMM-Newton}/RGS observations (\citealt{Bluestin+etal+2002}), and {\sl Suzaku}/XIS observations (\citealt{Brenneman11}).
In the UV regime, dynamic components are better constrained.
{\sl HST}/STIS and {\sl FUSE} spectra give the radial velocity of four WAs: 1350, 550, 725, and 1027 \kmps~(e.g., \citealt{Gabel+etal+2003b}).
By comparing photoionization models with individual lines, \cite{Gabel+etal+2005} constrained the $N_{\rm H}$ and $U$ for three of the four WAs.

We compare our results with theirs in Figure 7, but the results do not simply match with each other.
Since their studies use the systemic redshift of 0.009760 (\citealt{de+1991}) while we use 0.009815 of the NLR instead, their velocities are all supplemented by 17 \kmps.
WA 3 \& 4 match well two absorbers identified in the X-ray band by \cite{Netzer+etal+2003} in
velocity, ionization paramter, and conlumn density. On the other hand, The velocities of WA 1, 3, \& 4 also match two UV dynamic components measured by \cite{Gabel+etal+2003b}, but we do not detect their 1365 \kmps~component. In fact, \cite{Gabel+etal+2003a} found that this 1365 \kmps~component has a decrease in velocity of $\sim80$ \kmps~from 2000 to 2002 based on the STIS observations.
The authors speculated a possible mechanism for this deceleration that there is a directional shift in the motion of the WA with respect to our line of sight to the background emission sources, and deduced an upper limit time scale of 17 years that the WA would move outside the BLR.
The non-detection of this WA in our work supports this scenario.
Our WA 2 is not close to any previous identified one, though its properties are similar to WA 1.
It may just move inside our line of sight in a short time.
WA5 is likely at higher ionization state than previous detections, but still has similar column density.

\subsection{Pressure Balance}

\cite{Goncalves06}, \cite{Holczer+etal+2007}, and \cite{Goosmann+etal+2016}  constructed a continuous distribution of column density with ionization parameter, taking that all WAs are in pressure equilibrium.
We thus generate the thermal photoionization equilibrium curve (\citealt{Krolik+McKee+Tarter+1981}; known as S-curve) in Figure 8, but find not all five WAs are on the unstable vertical part which does not prefer the continuous distribution.
However, WA 3 \& 4 are likely in pressure balance.
They have similar velocities, column densities, and their covering factors are complemental.
It is reasonable to believe that they are the same absorbing gas, while different parts have slightly different ionization state.
WA 1 \& 2 are dropping out from the pressure balance region.
They have lower ionization parameters, lower column densities, lower covering factors, but higher velocities.
They may be smaller discrete clouds that are easily blown out from the inner region of the torus at different periods.

\begin{figure}
\centering
\includegraphics[angle=0,width=3.5in]{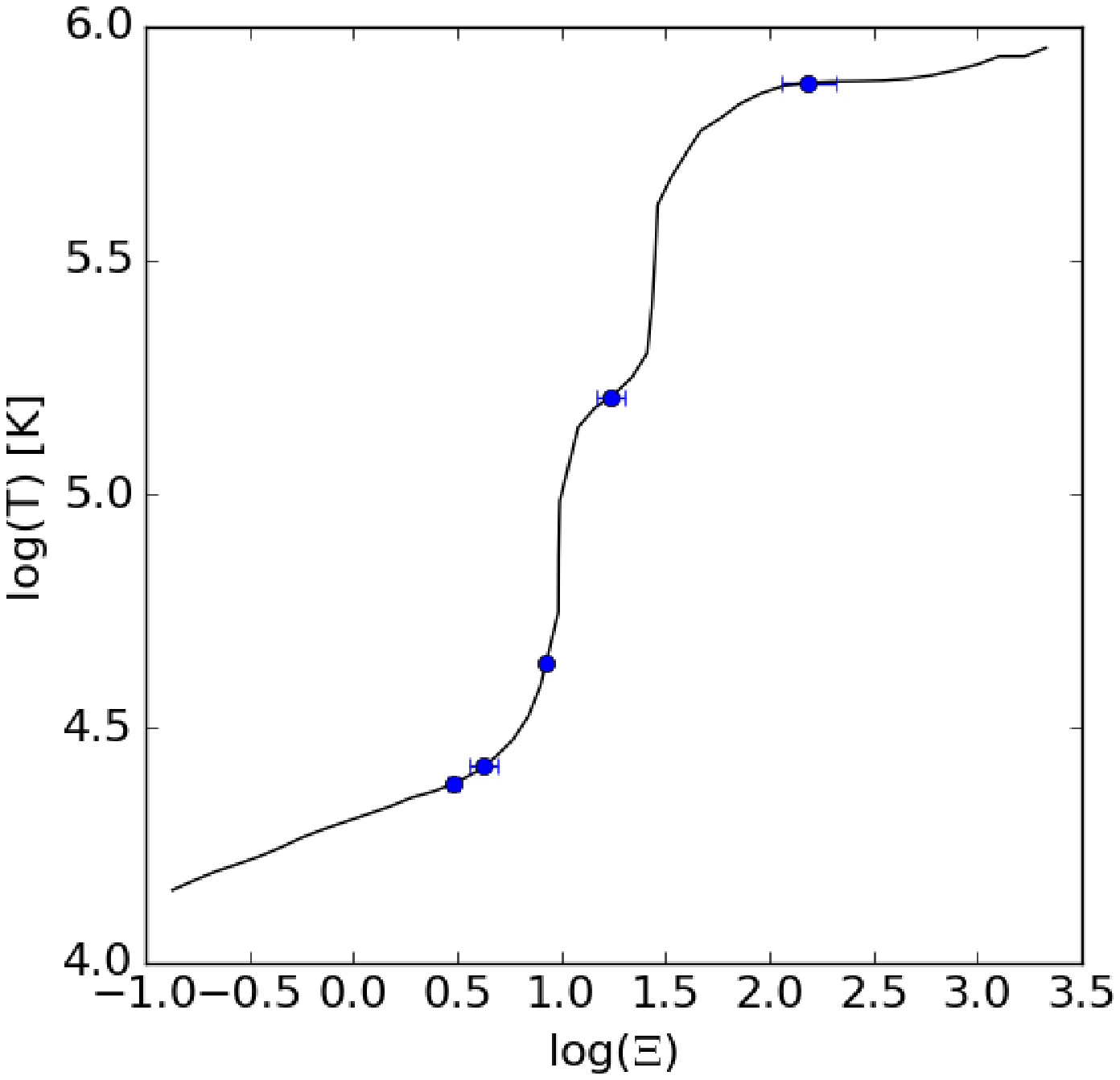}
\caption{The thermal stability curve, also called S-curve. The quantity $\Xi\equiv\xi/4\pi ckT$ is the pressure ionization parameter. The five blue points are for the five WAs.}
\end{figure}

\subsection{Mass Outflow Rate}
Bi-conical chimney is a natural geometry for WA outflows (\citealt{Dorodnitsyn+etal+2008}).
We use the following formula to estimate the mass loss rate, which is derived by \cite{Krongold+etal+2007}:
$\dot{M}_{\mathrm{out}}=0.8\pi m_{\mathrm{p}}N_{\mathrm{H}}v_{\mathrm{r}}Rf(\delta ,\phi )$, where $f(\delta ,\phi )$ is a factor that depends on the particular orientation
of the disk and the wind and, for all reasonable angles ($\delta>20^{\circ}$ and $\phi > 45^{\circ}$) it is of the order of unity.
$v_{r}$ is the line-of-sight outflow velocity.

According to the matching results in Section 5.1, the electron density of WA 4 is in the range of $1.2\times10^{2}$-$1.2\times10^{5}$ cm$^{-3}$ as derived by \cite{Gabel+etal+2005} and \cite{Netzer+etal+2003}.
Combining with the ionizing luminosity $L_{\rm ion} = 7.6\times10^{43}\,{\rm erg\,s^{-1}}$ and $\xi$ obtained, we estimate that WA 4 is 0.7-22.0 pc away from the nucleus. After taking the covering fraction into account, the mass outflow rate from WA 4 is 0.018-0.56 $M_{\odot}$ per year. Taking that WA 3 is likely at the same location of WA 4, our best-fit parameters of WA 3 indicate a mass
outflow rate of about 0.0035-1.1 $M_{\odot}$ per year. On the other hand, WA 1 and WA 2 may represent two blown-out clouds. We adopt a spherical-shape assumption for WA 1 and WA 2. Taking the BLR transverse size is of $1.9\times10^{16}$ cm (\citealt{Onken02}), and based on the best-fit values of corresponding covering factors, their sizes are estimated as $1.09\times10^{16}$ and $1.17\times10^{16}$ cm, respectively. Adopting those values as their thickness, the best-fit column densities of WA 1 and WA 2 indicate that their electron densities are about $8.34\times10^{3}$ and $3.18\times10^{3}$ cm$^{-3}$, respectively. The distances of 15.9 and 20.9 pc for WA 1 and WA 2 respectively can be derived. Their mass outflow rate are estimated following the same method as for WA 4, which are 0.010 and 0.005 $M_{\odot}$ per year. WA 5 is highly ionized and cannot be efficiently blowed away by the radiation pressure.
Its distance to the black hole should be larger than the inner radius of torus that is about 1 pc and less than other WAs which gives an upper limit of 15.9 pc. Based on these constraints, the range of mass outflow rate of WA 5 is about 0.15-2.4 $M_{\odot}$ per year. At last, the cumulated mass outflow rate of the five WAs is in the range of 0.22-4.1 $M_{\odot}$ per year. We can see that WAs disappeared and new WAs appeared during the past decade, so the mass outflow rate is statistical.

\section{Summary}
The bright Seyfert I galaxy NGC 3783 was observed by {\sl Chandra}/HETG and {\sl HST}/COS simultaneously in March 2013.
We perform a joint fit on these two band spectra to constrain the properties of WAs.

\begin{itemize}
\item We joint fit the two band spectra of NGC 3783 by considering the physical components of the local gas absorption, local dust extinction, AGN Comptonized corona emission, accretion disk black body emission, BLR emission, NLR emission, and the intrinsic WAs.
Finally five WAs can explain well all absorption lines in both the UV and X-ray band spectra.

\item The five WAs do not stay together at the pressure balance part of the S-curve. Two WAs are likely different parts of the same absorbing gas. While the other two WAs may be smaller discrete clouds that are blown out from the inner region of the torus at different periods.
The highest ionized WA has highest column density, which infers some tenuous but highly volume-occupied gas.

\item The total mass outflow rate of the five WAs is in the range of 0.22-4.1 $M_{\odot}$ per year.

\end{itemize}

\normalem
\begin{acknowledgements}
The work is supported by the National Natural Science Foundation of China under the grant 11203080 and  11573070. L. Ji is also supported by 100 talents program of the Chinese Academy of Sciences. We also thank for the help from Dr. Yangsen Yao on the UV data processing.

\end{acknowledgements}

%\bibliographystyle{raa}
%\bibliography{ms0095ref}

\end{document}